# Vision and Inertial Sensing Fusion for Human Action Recognition: A Review

Sharmin Majumder, *Student Member, IEEE*, Nasser Kehtarnavaz, *Fellow, IEEE*

*Abstract*—Human action recognition is used in many applications such as video surveillance, human–computer interaction, assistive living, and gaming. Many papers have appeared in the literature showing that the fusion of vision and inertial sensing improves recognition accuracies compared to the situations when each sensing modality is used individually. This paper provides a survey of the papers in which vision and inertial sensing are used simultaneously within a fusion framework in order to perform human action recognition. The surveyed papers are categorized in terms of fusion approaches, features, classifiers, as well as multimodality datasets considered. Challenges as well as possible future directions are also stated for deploying the fusion of these two sensing modalities under realistic conditions.

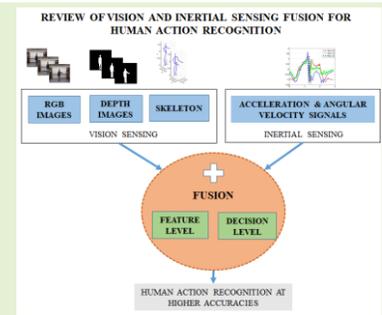

## I. Introduction

HUMAN action recognition means recognizing actions performed by humans based on action data captured by sensors such as RGB cameras, depth cameras, wearable inertial sensors, etc. In [1], [2], human activities have been categorized into four types: gestures, actions, interactions (with objects and others), and group activities. The focus of this survey paper is on human action/gesture recognition. Human action/gesture recognition has a wide range of applications including intelligent video surveillance [3], [4], home monitoring [5], human–machine interfacing, video storage and retrieval [6], assistive living, and assistant robots [7], [8], to name a few. It involves various research topics in computer vision including human detection in video, human pose estimation, human tracking, analysis and understanding of human activities. Research on human action recognition has made significant progress in the last decade and led to commercial products.

Action recognition can be achieved using different sensing modalities, most notably RGB video cameras, depth cameras, and wearable inertial sensors. In the literature, one sees two main sensing modalities: vision-based sensing and inertial-based sensing. A number of survey articles have already appeared involving a single modality sensing as well as multimodality sensing. The thrust of this survey paper is on the simultaneous utilization of vision and inertial sensing.

Early research on human action recognition was dominated by vision-based sensing, e.g. [9]–[16]. Most of these efforts used color and texture cues to obtain spatiotemporal volume-based features [17], joint trajectory features [18], [19], spatio-temporal interest points (STIP) [20], [21], spatio-temporal descriptors based on 3D gradients, motion-energy images (MEI), and motion history images (MHI) [22] using RGB video data. This sensing modality is found popular due to its wide availability and cost effectiveness. However, as noted in [23], there are still challenges in terms of difficulties posed by background clutter, partial occlusion, illumination changes, view-point variations, overlapping subjects, camera calibration, and biometric variations. In practice, a considerable amount of hardware resources are often needed to run computationally intensive video processing and computer vision algorithms. Furthermore, ambiguities are created because of 3D to 2D mapping associated with image or video data. As an example of video data, Fig.1(a) illustrates a sequence of video images for the *right-arm-swipe-to-the-left* action from the UTD-MHAD dataset [24].

The advancement of cost-effective RGB-D sensors (in particular, Kinect [25]) has led to the development of many depth-based human action recognition methods, e.g. [18], [26]–[32]. These methods are driven by the extra dimension (depth) generating 3D structural information of human actions. The extra depth dimension allows objects to be quickly segmented depthwise even in low illumination environments. In addition, the human skeleton information can be obtained from depth cameras. There are three types of commonly used depth cameras: stereo cameras (via triangulation), time-of-flight (TOF)-based cameras, and structured-light-based cameras. Structured-light and TOF-based depth cameras are affected by sunlight, limiting their utilization in outdoor environments. Although there are other sensors that can be used to measure depth, such as laser scanners, they are expensive and unsuitable for video surveillance and home monitoring applications. Recently, the newly developed Real Sense sensor has enabled analyzing distances as far as 10m. Compared to conventional RGB

The authors are with the Department of Electrical and Computer Engineering, University of Texas at Dallas, Richardson, TX 75080, USA (email: sharmin.majumder@utdallas.edu, kehtar@utdallas.edu)

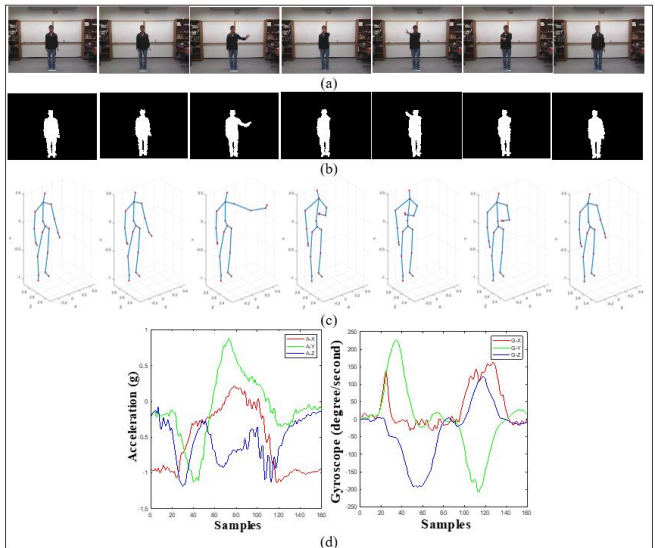

Fig. 1. An example of the multimodality data corresponding to the action right arm swipe to the left from UTD-MHAD dataset [24]: (a) the color images, (b) the depth images, (c) the skeleton joint frames, and (d) the inertial sensor data (3-axis acceleration and 3-axis gyroscope signals).

images generated by video cameras, depth images generated by depth cameras are shown to be insensitive to illumination changes and have led to achieving high recognition accuracies, in particular for indoor human action recognition. RGB-D sensors or cameras provide both the RGB video and depth information. An example sequence of depth images for the *right-arm-swipe-to-the-left* action from the UTD-MHAD dataset is illustrated in Fig. 1(b) and the corresponding skeleton frames are shown in Fig. 1(c). Depth data for the UTD-MHAD dataset were captured by a Kinect RGB-D camera.

In addition to vision-based sensors, wearable inertial sensors have been used for human action recognition allowing the recognition to be conducted beyond a limited field of view of a vision-based sensor. Inertial sensors contain accelerometers and gyroscopes providing acceleration and angular velocity signals which are used to recognize human actions, e.g., [33]–[39]. The advancements in lowering the energy consumption and increasing the computational power of inertial sensors have enabled long-term recordings and real-time computation by these devices. Similar to depth sensors, wearable inertial sensors provide 3D action data consisting of 3-axis accelerations from their accelerometers and 3-axis angular velocities from their gyroscopes. Wearable inertial sensors have their own limitations as well. For example, to fully capture the 3D motion associated with a human action, a single inertial sensor may not be adequate and it may be required to utilize multiple inertial sensors thus increasing the intrusiveness associated with wearing multiple sensors. An example of 3-axis acceleration and 3-axis gyroscope signals for the *right-arm-swipe-to-the-left* action from the UTD-MHAD dataset is shown in Fig. 1(d). The inertial signals for the actions reported in the UTD-MHAD dataset were recorded by a low-cost wireless inertial sensor described in [40].

Under real-life operating conditions, it is seen that no single sensing modality can cope with various situations that may occur in practice. One way to improve the performance of human action recognition is to use multimodal sensing. Multimodal sensing involves the fusion of two or more differing sensing modalities such as fusion of RGB and depth sensing, fusion of depth and inertial sensing, fusion of RGB and inertial sensing, and fusion of all these three sensing modalities. Multimodal sensing fusion has been shown to improve recognition accuracies compared to a single modality due to the complementary information provided by each sensing modality. For example, depth images provide a rich representation of global (or full body) movement attributes while inertial signals capture a rich representation of local movement attributes. In [41]–[51], it was shown that fusing depth and inertial sensing led to higher recognition accuracies compared to the accuracies when using each sensing modality individually. More specifically, in [52], [53], it was shown that higher recognition accuracies were reached when fusing RGB video and inertial sensing. Similarly, in [20], [54]–[58], higher recognition accuracies were obtained when fusing RGB video and depth sensing. In [59], [60], RGB, depth and inertial signals were simultaneously used to achieve higher recognition accuracies. Here, the thrust of this survey paper is placed on the fusion of vision and inertial sensing noting that both of these two sensing modalities can be deployed outdoors as well as indoors. Fig. 2 depicts both single and multimodal sensing modalities for human action recognition. The focus in this survey is on the vision and inertial fusion branch indicated by a box.

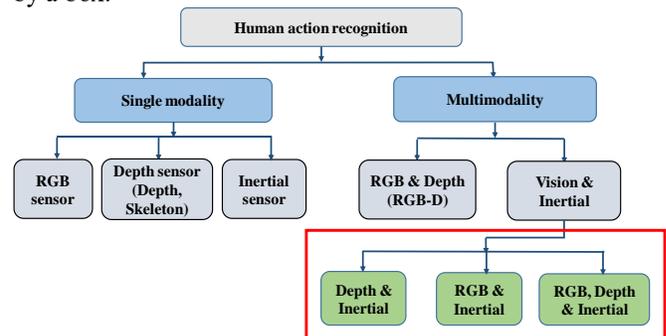

Fig. 2. Single and multimodality sensing for human action recognition; the box indicates the scope of this survey paper

There have already been several survey papers on human action recognition using RGB or video sensing alone [2], [61]–[69], depth sensing alone [23], [70]–[75], inertial sensing alone [76]–[79], RGB-D sensing [79]–[86], and fusion of depth and inertial sensing [88]. However, there has not yet been a survey paper focusing on the fusion of vision and inertial sensing modalities. For fusion of depth and inertial sensing modality, one survey paper appeared in 2015 [88]. Moreover, the existing survey papers on multimodal datasets mostly focus on the RGB-D multimodality [85], [89]. Recently, a survey paper has appeared on data fusion for action recognition [90], in which the data fusion techniques and classifiers for action recognition are reviewed. The existing survey papers on human action recognition are listed in Table I.

This paper first covers a survey of the papers on fusion of depth and inertial sensing as well as the papers on fusion of RGB and inertial sensing. Moreover, it reviews both handcrafted and deep learning features used. In addition, a list of 21 publicly available multimodal human action/gesture recognition datasets based on multimodality sensing data is provided. Basically, this survey paper in one place provides where the current technology of fusing vision and inertial sensing modalities stands for human action recognition.

The remainder of the paper is organized as follows: Section II covers an overview of fusion approaches used followed by a listing of publicly available multimodality action recognition datasets in Section III. Section IV covers the papers addressing the fusion of vision and inertial sensing modalities for human action recognition. In Section V, the challenges and possible future directions in the fusion of vision and inertial sensing are stated followed by the conclusion in Section VI.

## II. FUSION APPROACHES

The fusion of information from different sensing modalities has been conducted in different ways. In general, three fusion approaches are encountered as stated in [91]: (1) raw/data-level fusion, (2) feature-level fusion, and (3) score/decision-level fusion. The data-level fusion is applicable for homogeneous multi-sensors data (e.g., two or more RGB cameras or depth cameras, etc.). On the other hand, when there are two or more heterogeneous sensors, feature-level fusion or decision-level fusion techniques are normally applied. Fig. 3 provides illustrations of these fusion approaches.

### A. Raw Data-Level Fusion

In this fusion approach, raw data provided by sensors are combined before carrying out any processing. In other words, this fusion occurs at the data-level where incoming raw data from different homogenous sensors are combined. The feature extraction and classification processes are then performed on the combined data. The data-level fusion techniques typically involve estimation methods such as Kalman filtering as described in [91].

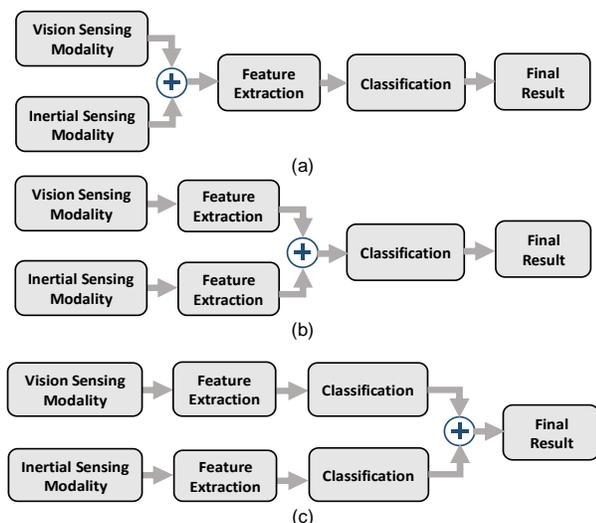

Fig. 3. Three fusion approaches in multimodal sensing: (a) Data-level fusion, b) Feature-level fusion, (c) Decision-level fusion

### B. Feature-Level Fusion

In this fusion approach, different features or feature vectors are fused to obtain a more comprehensive set or collection of features. The recognition process is carried out on the combined features. Multiple kernel learning [92], canonical correlation analysis (CCA) [93], Kernel canonical correlation analysis (KCCA) [94], discriminant correlation analysis (DCA), nonlinear common component analysis network [95], are the primary techniques that have been used to combine features from different sensors for human action recognition. For example, CCA maximizes the correlation of two groups of features in a projection space. In contrast, DCA not only maximizes the correlation of features across two feature sets, but also decorrelates features that belong to different classes within each feature set. Feature-level fusion involves carrying out fusion of features after raw data collection.

### C. Decision-Level Fusion

In this fusion approach, data collection, feature extraction, and classification are carried out for each individual sensing modality. A final decision is made by fusing the decisions from each sensing modality. Probabilistic methods (e.g., Naive Bayes Combination method (NBC) [96], Dempster-Shafer theory [97], [98]), linear opinion pool (LOP) [99], [100], logarithmic opinion pool (LOGP) rule [99], [100], majority voting rule [101], and weighted majority voting [101] have been used for decision-level fusion.

TABLE I
SURVEY PAPERS PREVIOUSLY APPEARED IN THE LITERATURE

| Data | Ref | Year | Tasks |
|---|---|---|---|
| Still images | [61] | 2014 | Action recognition |
| Video | [62] | 2014 | Action recognition, research and evaluation challenges |
| Video | [63] | 2017 | Deep learning based action recognition for single viewpoint, multi-viewpoint, and RGB-D datasets |
| Video | [64] | 2016 | Action recognition |
| Video | [13] | 2020 | Real-world challenges and solutions to vision-based sensor, overview of action recognition |
| Video | [102] | 2010 | Action representation, segmentation from input stream, and recognition |
| Still images, video | [2] | 2015 | Recognition of atomic actions, people interactions, human–object interactions, and group activities. |
| Still images, video | [65] | 2017 | Deep learning based human action recognition |
| RGB, depth | [84] | 2019 | Human action and interaction analysis |
| RGB, depth | [83] | 2019 | Action recognition methods, feature representation, human-object interaction |
| RGB, depth, hybrid | [82] | 2017 | Action and gesture recognition in image sequences |
| Skeleton | [71] | 2016 | Action recognition using 3D skeleton data |
| Skeleton | [23] | 2014 | Activity recognition for 3D skeleton data |
| Depth, skeleton | [70] | 2013 | Motion analysis |
| Depth, skeleton, hybrid | [72] | 2015 | Human action recognition |
| RGB, depth, skeleton | [103] | 2017 | Deep learning based human action recognition |
| RGB, depth, skeleton, hybrid | [81] | 2018 | Deep learning based human action recognition |
| Wearable sensors | [104] | 2016 | Activity detection and classification |
| Wearable sensors | [76] | 2013 | Activity recognition |
| Mobile, wearable sensors, video | [90] | 2019 | Data fusion techniques and multiple classifier systems for activity recognition |
| Depth and inertial fusion | [88] | 2015 | Action recognition using fusion of depth and inertial |

## III. PUBLIC DOMAIN MULTIMODAL DATASETS

The publicly available datasets for RGB, depth, skeleton, and inertial modalities reported in the existing papers have been collected by a motion capture system, structured-light cameras and time-of-flight cameras, and wearable inertial sensors containing accelerometers and gyroscopes. A list of 21 publicly available multimodal datasets are provided in Table II together with the performance of the recognition techniques used.

## IV. MULTIMODAL ACTION RECOGNITION BASED ON VISION AND INERTIAL SENSING

Research on multimodal human action recognition for vision and inertial sensing can be divided into three types: fusion of video and inertial sensing, fusion of depth and inertial, and fusion of video, depth, and inertial sensing. In what follows, a review of papers addressing these three fusion types is stated.

### A. Depth and Inertial Fusion

Fusion of depth and inertial sensing is proven to be effective for action recognition in indoor environments for applications such as smart homes, fall detection, and assistive living. The recent papers focusing on the combination of these two sensing modalities are covered next in terms of features, classifiers, and fusion approach used.

Depth data involves both depth image sequences and 3D skeleton joint positions, and inertial data involves 3-axis acceleration signals and 3-axis angular velocity signals. This subsection categorizes the papers depending on the data used for fusion: (a) depth images and inertial signals, (b) skeleton joint positions and inertial signals, and (c) depth images, skeleton joint positions, and inertial signals. In each category, handcrafted features and deep learning features are mentioned separately.

*a) Fusion of depth images and inertial signals*

In this subsection, the recent papers based on the fusion of depth images and inertial sensors are reviewed according to the type of features used by them: handcrafted features or deep learning features.

*Handcrafted features* - Simultaneous combination of depth images and inertial signals were considered in [48], [49], [51] to improve the accuracy of action recognition. Depth motion map (DMM) was considered in [48], [51], which was extracted from depth images to provide motion information associated with actions. 3D depth images acquired from a depth camera were first projected onto 2D orthogonal Cartesian planes to generate front-view, top-view and side-view depth images. DMMs were then obtained by adding difference frames generated from two consecutive frames for a complete duration of an action. In addition, four statistical features (mean, variance, standard deviation, and root-mean square) were extracted from 3-axis acceleration and 3-axis angular velocity signals of a wearable inertial sensor [48], [51].

In [49], Fast Fourier Transform (FFT) coefficients were extracted from acceleration signals, and histogram of oriented gradient (HOG) features were extracted from active, passive, and noise points of Motion Response Maps generated from depth images. Among them, active points represented the change of human body joints' positions during a movement and passive points described the background in depth images. Similarly, in [43], frequency domain features (Power Spectral Density (PSD) at a selected frequency band, the sum of the FFT coefficients, and the spectral entropy based on the power spectrum) and time domain features (Mean, Standard Deviation, Autocorrelation, Cross Correlation, Variance, RMS or Root Mean Square, MAD or Median Absolute Deviation, Inter-quadrature Range, Range, Minimum) were extracted from the inertial signals. PCA (Principle Component Analysis) was then used to select a combination of features with the largest possible variance.

The fall detection application is extensively studied using the fusion of depth and inertial sensing. In [105], acceleration and angular velocity signals from an inertial sensor, and the center of gravity of a moving person from a Kinect depth map were used in a fuzzy fusion inference module to detect falls. In [106], this approach was extended to obtain static poses and dynamic transitions. A thresholded sum vector of acceleration signals was used for an initial fall detection and then a fall event was verified through fuzzy inference on both depth and acceleration data. The depth map features used included the ratio of height to width of a person's bounding box in depth maps, a parameter expressing the height of a person's surrounding box in the current frame to the physical height of the person, the distance of a person's centroid to the floor, and the largest standard deviation from this centroid for the abscissa and the applicate, respectively. A person's pose was determined on the basis of depth maps, whereas the pose transitions were inferred using both depth maps and the accelerations acquired by a body worn inertial sensor.

Research has been carried out on energy efficient and reliable fall detection systems combining both inertial signals and Kinect depth images [107]–[110], where inertial signals were used to indicate a potential fall and depth images were used to authenticate the eventual fall. Depth maps were not processed frame by frame, rather were stored in a circular buffer. If the acceleration crossed a pre-selected threshold, the depth maps of a specified duration preceding the instant of fall were collected from a circular buffer and processed. In [107], [108], a nearest neighbor interpolation was used to fill the holes in depth maps and a person's blob was detected in depth maps. An SVM (Support Vector Machine) was used to acknowledge the presence of a person within the blob. Classifiers were trained on three features: the ratio of head-floor distance to the height of a person, the ratio expressing a person's area in the image to the area at an assumed distance from the camera, and the ratio of major length to major width of a blob representing a person. In [109], both the features in depth maps and point clouds were used to identify falls. In [110], depth difference maps were obtained from consecutive frames in depth images. Depth difference gradient maps were then generated from the difference maps and then their entropy was computed.

*Deep learning features* - Depending on the structure of the deep learning network used, the representative works are summarized below.

TABLE II
MULTIMODAL DATASETS FOR HUMAN ACTION RECOGNITION

| Dataset Name | Modality V D S I | #Actions, types | #Subject, #Samples | # views | Tasks | Ref | Modality/task | Accuracy |
|---|---|---|---|---|---|---|---|---|
| Berkeley-MHAD [117] | √ √ √ | 11 actions involving full body, upper extremities, and lower extremities | 12, 660 | Multi (4) | AR/PE/AS/SR | [51] | D,I/(AR) | 99.13 (SG) |
| | | | | | | [112] | D,I/(AR) | 99.8%(SS) |
| | | | | | | [50] | D,I/(AR) | 99.54%(SS) |
| UTD-MHAD [24] | √ √ √ √ | 27 actions | 8, 861 | Single | AR | [24] | D,I/AR | 79.1% (SG) |
| | | | | | | [45] | D,I/AR | 86.3% (SS) |
| | | | | | | [137] | D,I/(AR) | 89.2% (SG) |
| | | | | | | [41] | D,I/AR | 92.8% |
| | | | | | | [50] | D,I/(AS+AR) | 91.5 (SG), 97.2 (SS) |
| | | | | | | [138] | D,S,I/(AR) | 93.26% (SG) |
| | | | | | | [113] | S,I/(AR) | 95% (SG) |
| | | | | | | [52] | V,I/(AR) | 95.6% (SG) |
| | | | | | | [59] | V,S,I/(AR) | 97.9% |
| | | | | | | [111] | D,I/(AR) | 98.7% (SS) |
| | | | | | | [58] | V,D,I/(AR) | 98.3% (SG) |
| | | | | | | [112] | D,I/(AR) | 99.2% (SS) |
| | | | | | | [51] | V,I/(AD+AR) | 91.3%-smart TV gestures, 85.2% -sport actions (SG) |
| UTD-MHAD Multi-view dataset [24] | | 6 actions | 5, 900 | Multi (5) | AR | [48] | D,I/(AR) | 88.4% (SG) |
| UTD-CAD [45] | √ √ √ | 5 smart TV gesture actions | 5, 125 streams | Single | AR/AD | [41] | D,I/(AD+AR) | 97.0% |
| | | | | | | [45] | D,I/(AD+AR) | 96.2% (SS) |
| | | | | | | [42] | D,I/(AD+AR) | 90.2% (SS), 85.1% (SG) · F1 score |
| UTD-Dataset Transitions&Falls [44] | √ √ √ | Six transition movements and fall | 12, >=840 | Single | AR/FD | [44] | D,I/(AR) | 96.5% (SG), 97.6% (SS) |
| UTD-Dataset Continuous TransitionMovements [41] | √ √ | Six transition movements and fall | 5, 25 continuous streams | Single | AR/AD | [41] | D,I/(AD+AR) | F1 score: 90.9% |
| C-MHAD [118] | √ √ | 5 Smart TV gestures and 7 transition movements | 12, 240 videos clips | Single | AR/AD/FD | [109] | V,I/(AD+AR) | 81.8% for smart TV gestures, 82.3% for transition movements (SG) |
| Kinect 2D [139] | √ √ √ | 10 actions | 6,300 | Single | AR | [47] | D,S,I/(AR) | 93.7% (SG), 99.4% (SS) |
| | | | | | | [111] | D,I/(AR) | 99.8% (SS) |
| | | | | | | [112] | D,I/(AR) | 99.8% (SS) |
| CAS-YNU-MHAD [140] | √ √ √ | 10 actions | 10, >=1085 | - | AR | [140] | D,I/(AR) | 96.76% |
| | | | | | | [49] | D,I/(AR) | 96.91% (SG) |
| 50 salads dataset [141] | √ √ | 2 mixed salads, each activity is associated with one of these activities: prepare dressing, cut and mix ingredients, and serve salad. | 25, 966 | Single | AR/AD | [141] | D,I/(AR/AD) | >=49% for all activities |
| ChAirGest [142] | √ √ √ √ | 10 gestures | 10, 1200 | Single | AR/AD | [142] | D,I/(AR) | F1 score: 86% |
| TST fall detection [143] | √ √ √ | 4 daily activities and 4 falls | 11, 264 | Single | AR/FD | [143] | D,I/(AR) | 99% |
| URFD dataset [144] [108] | √ √ √ | 2 falls and daily activities | 5, 30 falls and 40 daily activities | Multi (2) | AR/FD | [144] | D,I/(FD) | 98.33% |
| | | | | | | [109] | D,I/(FD) | 95.71 |
| | | | | | | [106] | D,I/(FD) | 97.14% |
| Huawei/3DLife dataset [145] | √ √ √ | 22 actions involving upper human body, training exercises, sport activities | 17, 3740 | Multi (5) | AR/SR | [145] | D,I/(AR) | Side-view: 82.8% Front-view: 88.1% |
| SLD [146] | √ √ √ | 10 sign language dataset | 5,250 | Single | AR | - | - | - |
| UP-Fall Detection dataset [147] | √ √ | 11 (6 daily activities+5 falls) | 17, >500 | Multi (2) | AR/FD | - | - | - |
| NCTU-MFD [127] | √ √ √ √ | 10 fitness activities | 10, 1200 | Multi (3) | AR | - | - | - |
| CMU-MMAC [132] | √ √ | 5 different recipes (complex activities) | 5, 30 min video for each subject | Multi (6) | AR/AD | [120] | V,I/(AR) | F1 score: 58.4% |
| ADL dataset [120] | √ √ | 6 complex activities | 2, - | Single | OD/AR | [119] | V,I/(AR) | F1 score: 45.8% |
| | | | | | | [120] | V,I/(AR) | F1 score: 79.3% |
| Multimodal Egocentric Activity dataset [123] | √ √ | 20 life-logging activities (Ambulation, daily activities, office work, and exercise) | >=2, >=200 | Single | AR/AD | [123] | V,I/(AR) | 83.71% |
| RealWorld (HAR) [148] | √ √ | 8 daily activities | 15, 10 min video for each subject & each action | Single | AR/AD/FD | - | - | - |

V-Video, D-Depth, S-Skeleton, I-Inertia, AR-Action Recognition, PE-Pose Estimation, SR-Scene Reconstruction, AD-Action Detection, AS-Action Segmentation, FD-Fall Detection, OD-Object Detection, SS-Subject-Specific, SG-Subject-Generic

Two 2D CNN architectures were considered in [111], one for each sensing modality. For depth modality features, sequential front-view depth images were fed into a pre-trained AlexNet CNN. A 2D signal image was also generated by row-by-row stacking of six inertial signals (3-axis acceleration signals and 3-axis angular velocity signals) and fed into a second CNN for inertial modality feature extraction. A shared feature layer was generated by fusing the features extracted from the two CNNs. Similar signal images were considered in [112] for a multimodal fusion approach. For the depth modality, two AlexNets and for the inertial modality, two CNNs were used in this paper. Sequential front-view images generated from front-view depth images were inputted to one AlexNet and the Prewitt filtered depth images were used as inputs to another AlexNet. For the inertial modality, the generated signal image was fed into one CNN and the Prewitt filtered signal image was fed into another CNN. In [44], six transition movements as well as falls were recognized by applying a two-stream CNN. Weighted depth images were generated from depth images and inputted to one CNN. Inertial signal images generated using eighteen heuristic orientation invariant transformation signals computed from the overall acceleration and overall angular velocity signals were inputted to another CNN.

In [41], a two-stream CNN, including one 3D CNN and one 2D CNN, were used for the depth modality and one CNN-LSTM with handcrafted features was used for the inertial modality. Sequences of depth images were fed into the 3D CNN and weighted DMMs were fed into the 2D CNN. Signal images were generated from the inertial signals using 3-axis acceleration signals, 3-axis angular velocity signals, overall acceleration signals, and overall angular velocity signals, and fed into a CNN-LSTM network. Handcrafted statistical features (mean, variance, standard deviation, root mean square, median, minimum and maximum) were also extracted from the inertial signals.

*a)* *Fusion of skeleton and inertial signals*

Simultaneous utilization of skeleton and inertial signals were considered in [42], [45], [46], [113]–[116] for action recognition. In [46], the raw signals of the 3-axis acceleration signals, 3-axis angular velocity signals, and 3D skeleton joint positions were normalized to classify five hand gestures. In [42], [45], normalized relative orientations (NROs) were computed for all the skeleton joints by considering each joint position and its rotating joint position. In [42], a potential energy function was computed for skeleton joints using NROs and then the potential energy difference between two successive frames were used to separate motion and pause segments in continuous action streams. Whenever a motion segment ended, the likelihood probabilities of the motion segments were obtained for each motion cluster difference based on which actions were recognized. The acceleration and gyroscope signals were then used to remove false positives. In [45], NROs from the skeleton data and the mean, variance, standard deviation and root mean square features, similar to [51], for the acceleration and angular velocity signals were used for action recognition.

In [113], 20 skeleton joint positions were stacked column-wise. Similarly, the acceleration and gyroscope signals were also stacked. To reduce temporal variations, all the signals were made of the same size with respect to the entries using bicubic interpolation. In [114], the spine-base joint position (located at the base of the spine) and the distance of the joint from the floor were considered. 3-axis acceleration signals were converted into gravity accelerations to compute an acceleration magnitude. In [116], a system for monitoring various dining activities of post-stroke patients was presented using a Kinect camera and accelerometers. 3D-trajectories of a subject's head, left and right hand were estimated while carrying out eating and drinking tasks.

*b)* *Fusion of depth, skeleton, and inertial signals:*

Combinations of depth images, skeleton joint positons, and inertial signals were used in [48], [47], [50]. In [47], the DMMs of depth images and the statistical features similar to [51] were extracted from skeleton joint positions and inertial signals. Similar statistical features for inertial signals and DMMs were also considered in [48], [50]. In [48], a view invariant action recognition was conducted by considering five viewing angles (front, left 45, left 90, right 45, and right 90). A viewing angle was estimated first and then the action was recognized for that particular view. In [50], only front-view DMM was considered to keep the computation time low noting that the use of the other two DMMs did not have much impact on the accuracy.

*Classification and Fusion* - Three types of fusion approaches were stated earlier in Section III. The feature-level and decision-level fusion are the approaches normally used for human action recognition. Feature-level fusion was considered in [43], [46], [48], [49], [51], [111]. In [46], [48], [51], features were simply concatenated to form a fused feature vector, e.g., nine-dimensional vector in [46]. To take into consideration temporal signal sequences, a left-right HMM topology was adopted in [46] to recognize five hand gestures. In [111], a shared feature layer was used after the multimodal fusion and then a support vector machine or a softmax classifier was used to recognize the actions based on the combined features. An ensemble classifier was utilized in [43].

In several papers, it is reported that the decision-level fusion is found more effective than the feature-level fusion for the datasets examined. In [47], 3 CRCs (Collaborative Representative Classifier) were used to classify the actions individually followed by a decision-level fusion based on logarithmic opinion pool (LOGP). In [50], 2 CRCs were used together with LOGP for a decision-level fusion. In [45], a support vector data descriptor (SVDD) and a CRC classifier were used for depth and inertial sensing modalities, respectively, followed by a decision-level fusion. In [113] a neural network (NN) was used for each modality and a LOGP based decision-level fusion was considered. In [44], the decisions from two CNNs corresponding to the two modalities were combined by a decision-level fusion. In [115], a NN was trained for each activity to distinguish all the samples belonging to one activity from a randomly chosen set of samples belonging to all other activities. A set of binary classifiers based on feedforward neural networks was combined to achieve action recognition. In [107], [108], a k-NN classifier and a linear SVM classifier were used to check whether a person was lying on the floor using depth images. In [109], a k-NN classifier was used to implement an exemplar-based fall detector. In [110], recognition was accomplished using a sparse representation-based classifier.

In [41], a two-stream CNN, including one 3D CNN and one 2D CNN, were used for the depth modality and one CNN-LSTM with handcrafted features was used for the inertial modality. Sequences of depth images were fed into the 3D CNN and weighted DMMs were fed into the 2D CNN. Signal images were generated from the inertial signals using 3-axis acceleration signals, 3-axis angular velocity signals, overall acceleration signals, and overall angular velocity signals, and fed into a CNN-LSTM network. Handcrafted statistical features (mean, variance, standard deviation, root mean square, median, minimum and maximum) were also extracted from the inertial signals.

### c) Fusion of skeleton and inertial signals

Simultaneous utilization of skeleton and inertial signals were considered in [42], [45], [46], [113]–[116] for action recognition. In [46], the raw signals of the 3-axis acceleration signals, 3-axis angular velocity signals, and 3D skeleton joint positions were normalized to classify five hand gestures. In [42], [45], normalized relative orientations (NROs) were computed for all the skeleton joints by considering each joint position and its rotating joint position. In [42], a potential energy function was computed for skeleton joints using NROs and then the potential energy difference between two successive frames were used to separate motion and pause segments in continuous action streams. Whenever a motion segment ended, the likelihood probabilities of the motion segments were obtained for each motion cluster difference based on which actions were recognized. The acceleration and gyroscope signals were then used to remove false positives. In [45], NROs from the skeleton data and the mean, variance, standard deviation and root mean square features, similar to [51], for the acceleration and angular velocity signals were used for action recognition.

In [113], 20 skeleton joint positions were stacked column-wise. Similarly, the acceleration and gyroscope signals were also stacked. To reduce temporal variations, all the signals were made of the same size with respect to the entries using bicubic interpolation. In [114], the spine-base joint position (located at the base of the spine) and the distance of the joint from the floor were considered. 3-axis acceleration signals were converted into gravity accelerations to compute an acceleration magnitude. In [116], a system for monitoring various dining activities of post-stroke patients was presented using a Kinect camera and accelerometers. 3D-trajectories of a subject's head, left and right hand were estimated while carrying out eating and drinking tasks.

### d) Fusion of depth, skeleton, and inertial signals:

Combinations of depth images, skeleton joint positons, and inertial signals were used in [48], [47], [50]. In [47], the DMMs of depth images and the statistical features similar to [51] were extracted from skeleton joint positions and inertial signals. Similar statistical features for inertial signals and DMMs were also considered in [48], [50]. In [48], a view invariant action recognition was conducted by considering five viewing angles (front, left 45, left 90, right 45, and right 90). A viewing angle was estimated first and then the action was recognized for that particular view. In [50], only front-view DMM was considered to keep the computation time low noting that the use of the other two DMMs did not have much impact on the accuracy.

*Classification and Fusion* - Three types of fusion approaches were stated earlier in Section III. The feature-level and decision-level fusion are the approaches normally used for human action recognition. Feature-level fusion was considered in [43], [46], [48], [49], [51], [111]. In [46], [48], [51], features were simply concatenated to form a fused feature vector, e.g., nine-dimensional vector in [46]. To take into consideration temporal signal sequences, a left-right HMM topology was adopted in [46] to recognize five hand gestures. In [111], a shared feature layer was used after the multimodal fusion and then an SVM or a softmax classifier was used to recognize the actions based on the combined features. An ensemble classifier was utilized in [43].

In several papers, it is reported that the decision-level fusion is found more effective than the feature-level fusion for the datasets examined. In [47], 3 CRCs (Collaborative Representative Classifier) were used to classify the actions individually followed by a decision-level fusion based on logarithmic opinion pool (LOGP). In [50], 2 CRCs were used together with LOGP for a decision-level fusion. In [45], a support vector data descriptor (SVDD) and a CRC classifier were used for depth and inertial sensing modalities, respectively, followed by a decision-level fusion. In [113] a neural network (NN) was used for each modality and a LOGP based decision-level fusion was considered. In [44], the decisions from two CNNs corresponding to the two modalities were combined by a decision-level fusion. In [115], a NN was trained for each activity to distinguish all the samples belonging to one activity from a randomly chosen set of samples belonging to all other activities. A set of binary classifiers based on feedforward neural networks was combined to achieve action recognition. In [107], [108], a k-NN classifier and a linear SVM classifier were used to check whether a person was lying on the floor using depth images. In [109], a k-NN classifier was used to implement an exemplar-based fall detector. In [110], recognition was accomplished using a sparse representation-based classifier.

Both feature-level and decision-level fusion were examined in [51]. The PCA dimensionality reduction was applied to the fused feature vector. For feature-level fusion, a sparse representative classifier (SRC) [117] was utilized to classify the combined features and for decision-level fusion, two CRC classifiers were utilized to classify the actions separately for each modality. In [42], a Maximum Entropy Markov Model (MEMM) classifier was used to detect and recognize actions from continuous action streams using skeleton data and a CRC classifier was used to improve the recognition outcome using inertial signals. In [41], [112], a combination of feature-level and decision-level fusion were employed for action recognition. In [41], the features from a 3D CNN and a 2D CNN were combined in the depth modality and fed into a softmax layer classifier. In the inertial modality, another softmax classifier makes the decision based on the combined features from a CNN-LSTM network together with the handcrafted features. Detection and recognition are performed for each of the two sensing modalities in parallel followed by a decision-level fusion. In [112], the features obtained from two CNNs were combined for the inertial modality and the features from two AlexNets were combined for the depth modality. Three fusion frameworks were investigated including deep multistage feature fusion, computationally efficient fusion, and deep hybrid fusion. The deep hybrid fusion framework achieved the highest accuracy among these three frameworks for the datasets considered.

### B. Video and Inertial Fusion

One 3D CNN for RGB video and one 2D CNN for inertial signal images were used in [52], [53], [118]. Similar to [41], inertial signal images were generated from 3-axis acceleration signals, overall acceleration signal, 3-axis angular velocity signals, and overall angular velocity signal by row-wise stacking. In [119]–[121], an action recognition approach was introduced improving the motion-based action recognition with egocentric vision. In [120], [121], inertial data were collected from a smart-watch and video data were collected from a pair of smart-glasses to recognize actions. The inertial data were used to characterize the forearm movement pattern whereas the egocentric video data was used to characterize objects. For object detection, a ResNet FPN (Feature Pyramid Network) model [122] was considered and the features from both the time domain (mean, median, standard deviation, variance, inter quantile range, MAD, kurtosis, correlation coefficient, gravity, orientation, entropy) and the frequency domain (energy, entropy, mean DC) were used. Interactions with objects were achieved by looking at the objects overlapped with a detected hand using a pre-trained neural network.

In [123], a fusion method was developed for inertial and video sensing using a Fisher Kernel framework. A generative model was considered followed by the Fisher Kernel to obtain a multimodal feature vector for a discriminative classifier. The Fisher kernel combined the strengths of generative and discriminative approaches. A trajectory-based approach was used for both sensing modalities. Dense trajectories were first obtained using optical flow fields and then MBH (motion boundary histogram) descriptors were computed from the trajectories. For the inertial modality, time-series inertial signals were converted into trajectory-like features and a temporal order was introduced to enhance them.

A combination of video and inertial data were used for fall and staggering detection in [124], and for nocturnal epileptic seizure detection in [125]. Since the volume of the recorded video data was very large, a smartphone inertial sensing triggering mechanism was developed in [124] to select and process only the relevant video data. The initial motion orientation was further mapped into image coordinates to guide the direction of human body tracking. The spatiotemporal information was used to retrieve corresponding video clips. The variation of 3D acceleration provided hints regarding the motion direction. Ten-dimensional features consisting of mean, variance, number of zero-crossing points, auto- and cross-correlation of vertical and horizontal acceleration were extracted from the acceleration signals. After getting triggered, the inertial sensor generated a 3D velocity vector which was then projected onto the 2D image plane and used as the initial velocity. A human detection algorithm [126] was applied to detect the human body as well as the torso position. Dense optical flow features were extracted from the tracked images of a candidate sequence.

To detect nocturnal epileptic seizure, in [125], the collected data were first segmented into normal or epileptic movements. Segmentation was achieved based on the accelerometer signals using the time and frequency domain features of mean amplitude, standard deviation, max amplitude of resultant, correlation, kurtosis, RMS, skewness, peak frequency, energy, spectral edge frequency, power in sub-bands and the normalized power in sub-bands. Normal movements were modeled based on the probability density function (pdf) of the features. The movements that were found to be outliers with respect to the pdf in the testing phase were taken to be epileptic movements. For classification of seizures, spatio-temporal interest points were first extracted from video data generating multiple spatio-temporal cuboids. Then, the histograms of the optical flow (HoF) features were extracted from the cuboids.

*Classification and Fusion* - In [52], [118], [125], a decision-level fusion approach was considered. In [125], video sequences were classified using a bag-of features approach. The histograms of video sequences were used as the input to a support vector machine (SVM) classifier with the output giving a probability of the epileptic seizure. The outputs from the segmentation and classification algorithms were used to serve as the input to a LDA classifier to achieve fusion.

In [53], [120], [121], both feature-level fusion and decision-level fusion were examined with the decision-level fusion achieving higher accuracy. For the feature-level fusion, after combining the features from the two modalities, a softmax classifier was used to make the final decision. For the decision-level fusion, two softmax classifiers were used to make two decision separately and then a fused decision was done via the Borda count method. In [120], [121], a Random Forest classifier was used for the acceleration data and Logistic Regression for the vision data. In [124], a ten-dimensional feature vector was first formed using statistical features from the acceleration signals. A one-versus-one SVM was trained to perform the classification. The final decision was reached by the fusion of a confidence score from the SVM classifier and the behavior likelihood distribution obtained from a module using the inertial sensor data.

### C. RGB, Depth, and Inertial Fusion

This section reviews the works reported in [59] [60] on fusion of RGB video, depth, and inertial sensing. 3D HOG features from RGB and depth videos, and three time domain features from acceleration and gyroscope signals were considered in [59]. A codebook was constructed to generate a bag-of-words based on the histogram of features followed by a feature-level fusion. KNN and SVM classifiers were utilized and it was found that the KNN classifier worked better than the SVM classifier. In [60], a 1D CNN for 3-axis gyroscope signals, a 2D CNN for RGB video, and an RNN for skeleton joints sequences were used. Stacked dense flow difference images (SDFDI) were generated from the RGB video and fed into the 2D CNN. SDFDI was generated by taking optical flow differences between consecutive frames and stacked together for all the frames of an action. SDFDI was capable of capturing video spatio-temporal information. In the RNN, two bidirectional gated units (BiGRU) were applied in both forward and backward directions to process the input skeletons. This bi-directional GRU approach performed better than the traditional one-directional LSTM models. The classified outputs from three heterogeneous networks (1D-CNN, 2D-CNN and BiGRU) were combined via a decision-level fusion.

## V. Challenges and possible future directions

Although there are challenges in any sensing modality system towards performing human action recognition under realistic conditions, there are some challenges that are unique to multimodality fusion of vision and inertial sensing which are stated in this section together with possible future directions to deploy multimodality fusion of vision and inertial sensing for human action recognition.

### A. Challenges

#### 1) Data synchronization

A main challenge in multimodal sensing systems is data synchronization. Data synchronization means synchronizing the time samples of actions acquired from different sensors. If the data associated with an action from the two sensing modalities are not synchronized in time, the recognition accuracy will get adversely impacted. The publicly available multimodal datasets normally provide segmented actions data that are synchronized across the sensing modalities.

In [24], the start and the end of an action were synchronized by using the timestamps of the depth images to serve as references due to variations of the frame rate of the Kinect camera and the sampling rate of the wearable inertial sensor. Similarly, in [127], the start and the end of an action were synchronized by using the date/time of the Kinect camera (either RGB images, depth images or skeleton joints) as references. In [128], the data from a Kinect camera and a wearable inertial sensor were collected using C++ codes. The skeleton joint positions at each frame were read as binary files and each sample of acceleration and gyroscope signals was stored in a text file. A Matlab code was used to start searching the first sample from the inertial data and the first frame of the skeleton data in parallel. There was a time stamp index value associated with each frame of the skeleton joint positions which provided the corresponding sample number of the inertial signals.

In [46], a time synchronization approach was presented by correlating the closest inertial sample to the depth frame based on the system timestamps. In [117], data synchronization among different modality sensors was achieved by using the UNIX operating system timestamps. In [129], a time synchronization method was implemented by estimating the total delay occurring in the link between a video camera and a computer. This approach was later used in [114] to synchronize depth and inertial data for the fall detection application. In [130] [131], a calibration was performed to align the RGB and depth images acquired from a Kinect camera.

In [129], the transmission and exposure times of the frames captured by a Kinect camera was utilized to synchronize the RGB-D sensor with two inertial measurement units (IMUs). An acquisition software allowed to simultaneously capture data from the Kinect camera connected via a USB cable, and from the accelerometers via a Bluetooth link to the same PC. The software applied a timestamp when each packet, or frame, arrived at the PC. The synchronization was realized by exploiting these timestamps, taking into account the transmission times of Kinect frames and any possible delays caused by the Bluetooth protocol.

In [132], synchronization among video, audio, motion capture, and inertial signals was achieved by combining two different protocols: MultiSync and Network Time Protocol (NTP). The MultiSync software is designed to synchronize image acquisition of multiple compatible Point Grey cameras across different IEEE-1394b buses on the same computer and across separate buses on multiple computers. Moreover, it records the system timestamp in order to be able to synchronize cameras with other devices. NTP is a protocol for synchronizing the clocks of computer systems over packet-switched, variable-latency data networks.

#### 2) Online processing

Another main challenge in human action recognition when using more than one sensing modality is online processing. Online processing involves ability to detect actions in continuous data streams in an on-the-fly manner, referred to as continuous action recognition. The great majority of the papers that have appeared in the literature on human action recognition have studied action signals (e.g., RGB videos, depth videos, inertial sensor signals) that have already been segmented manually or by visual inspection. Compared to segmented action recognition, continuous action recognition is more challenging since the action segmentation and detection need to be performed first in real-time noting that actions can occur at different execution speeds. For continuous action recognition, three steps are needed: segmentation, detection, and recognition. Segmentation separates actions from non-activity in a continuous action stream; detection separates the actions of interest in a particular application from actions of non-interest; and finally recognition classifies the detected actions of interest. Action segmentation methods were discussed in [102], [133] when using vision sensing individually and in [134], [135] when using inertial sensing individually.

For multimodal depth and inertial sensing fusion, both depth and inertial data were considered in [41], [50] to achieve action segmentation separately for each modality. Assuming each action began with a static posture and ended with a static posture lasting at least one second, the variances of the skeleton joint positions and the accelerations within a moving window were used to determine the start and the end of actions. In [50], 3D joint positons and acceleration signals were captured first for static postures. Action segmentation for the depth modality was achieved by taking the distance between the joint positions of each frame and a static posture. Similarly, for the inertial modality, magnitude differences between each sample and a static posture were used to achieve action segmentation. In [41], for the depth modality, centroid differences in depth images were computed between successive frames and the frames above a specified threshold value were assigned as segmented actions. Similarly, for the inertial modality, angular velocity differences between two successive samples were computed and the samples above a specified threshold value were assigned as segmented actions. In [42], [45], detection and recognition of smart TV gestures from continuous action streams were performed. In [45], potential energy from skeleton joints positions (computed from NROs) and acceleration differences (computed by taking differences of overall acceleration and reference acceleration) were used for action detection. In [136], two Gaussian models, one for rest positions and the other for non-rest positions, were used to classify the observations (a combination of hand positions from a depth camera and inertial signals from an inertial sensor) into a rest or a non-rest position during testing of hand gestures. A sequence of continuous observations from non-rest positions longer than a quarter second was then considered to denote a gesture.

In [52], [118], a real-time action recognition system was developed based on the video and inertial modalities. Segmentation, detection and recognition were done for both of the modalities separately. For the inertial signals, segmentation was done using the acceleration difference of each frame and a reference frame. This corresponded to an overall acceleration difference for an action sequence containing several actions of interest among actions of non-interest. For the video modality, the grayscale intensity difference between two consecutive frames was considered which corresponded to the mean brightness difference signal. Difference signals less than 1% of the signal maximum were considered to be non-activity.

### B. Possible Future Directions

In order to improve the accuracy and robustness in multimodal action recognition systems, certain topics need further investigation. A list of possible future research topics or directions is stated below.

• *Multimodal action recognition datasets:* There are few publicly available datasets for multimodal sensing involving vision and inertial data. The existing multimodal datasets mainly involve only RGB and depth data with no synchronized inertial data. The availability of multimodal vision and inertial datasets can provide a uniform framework for comparison of fusion techniques as related to vision and inertial sensing

• *Continuous action recognition:* The publicly available multimodal action recognition datasets mainly provide segmented action data. In real-world scenarios for a particular application, the actions of interest occur in continuous manner among many other actions of non-interest. Therefore, it is required to perform action segmentation in an on-the-fly manner or in real-time. In other words, practical cost-effective systems need to address computationally efficient solutions for segmentation, detection, and recognition. Any computationally intensive solution or algorithm would hinder the practical utilization of an action recognition system in real-time.

• *Complex activities:* Most existing methods address simple actions, such as walking, running, and sitting, among others. In practice, these simple actions are part of more complex activities (e.g., playing tennis). Complex activities are composed of a sequence of several actions performed in a particular order. A future research direction can involve the design of systems for recognizing complex activities that are made up of a series of actions.

• *Action prediction:* Action recognition has proven effective in many applications such as health monitoring, human-robot interaction, and human computer interaction. However, in some applications such as fall detection, it is important to be able to predict the likelihood of an action event for the purpose of taking evasive countermeasures. The prediction of a particular action is another area of possible future direction noting that not much work has been reported related to action prediction.

• *User-specific or personalized model:* People perform actions differently and at different speeds. In other words, the models designed based on a dataset of a few subjects do not generalize well to all people. One possible future direction is to start with a subject-generic model and tune that model for a specific user in a subject-specific manner. This allows a subject-generic model to be personalized.

## VI. Conclusion

This paper has presented a survey of multimodal fusion of vision and inertial sensing for human action recognition. A comprehensive collection of papers have been provided summarizing the fusion works that have been conducted so far in the literature as related to fusing vision and inertial sensing. Challenges as well as possible future directions are also stated for deploying the fusion of these two sensing modalities under realistic conditions.

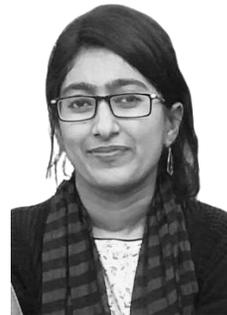

**Sharmin Majumder** (S'17) received the BS and MS degrees in Electrical and Electronic Engineering (EEE) from the Chittagong University of Engineering and Technology (CUET), Chittagong, Bangladesh, in 2013 and 2019, respectively. She is currently pursuing the PhD degree in Electrical Engineering at the University of Texas at Dallas, Richardson, TX. She also holds a faculty position with the Department of EEE at CUET. Her research interests include signal and image processing, computer vision, pattern recognition, and machine learning.

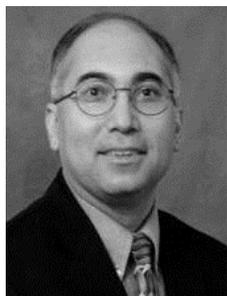

**Nasser Kehtarnavaz** (S'82–M'86–SM'92–F'12) is an Erik Jonsson Distinguished Professor with the Department of Electrical and Computer Engineering and the Director of the Embedded Machine Learning Laboratory at the University of Texas at Dallas, Richardson, TX. His research interests include signal and image processing, machine learning and deep learning, and real-time implementation on embedded processors. He has authored or co-authored 10 books and more than 400 journal papers, conference papers, patents, manuals, and editorials in these areas. He is a Fellow of SPIE, a licensed Professional Engineer, and Editor-in-Chief of *Journal of Real-Time Image Processing*.